\documentclass[letterpaper,aps, twocolumn]{revtex4}
\usepackage{amsmath}
\usepackage{amsfonts}
\usepackage{amssymb}
\usepackage{graphicx}
\bibstyle{apsrev}
\usepackage[utf8]{inputenc}

\newcommand\scalemath[2]{\scalebox{#1}{\mbox{\ensuremath{\displaystyle #2}}}}
\emergencystretch=\maxdimen
\begin{document}

\title{Dirac's reduction of linearized gravity in $N > 2$ dimensions revisited}
\author{R. Gaitan$^\dag$ and E. Schiappucci$^\dag$}
\affiliation{ $\dag$ Departamento de F\'{\i}sica, Facultad Experimental de Ciencias y
Tecnolog\'{\i}a, \\Universidad de Carabobo, Valencia - Venezuela}

\begin{abstract}

We perform a brief review on Dirac's procedure applied to the well
known Einstein's linearized gravity in $N > 2$ dimensions.
Considering it as a gauge theory and therefore the manifestation
of second class constraints in analogy with the electromagnetic
case, focussing our interest in the Coulomb's gauge. We also check
the consistency with the Maskawa-Nakajima reduction procedure and
end with some remarks on both procedures.

\end{abstract}

\maketitle

\section{Introduction}

For a long time there has been great interest on the procedure to
quantize gravity in a perturbative regime using known techniques
coming from Quantum Field Theory. In the Fock's space formalism,
for example, it is confirmed that graviton's spin is 2 thanks to
the representation of the Poincar\'e algebra via the
creation-annihilation operators\cite{Gupta}. Later, a series of
studies began in the context of Hamiltonian formalism with the
goal of reducing the degrees of freedom of Einstein's gravity by
the imposition of constraints that don't change the equations of
motion, even though they change the Lagrangian density and we must
abandon covariance\cite{Dirac}.

There are other perspectives to engage the degrees of freedom
reduction of gravity in the Hamiltonian formalism and the Dirac's
constraints analysis. The light-cone\cite{Lee} or the
null-plane\cite{EvensK} coordinates fixings which not necessarily
means a priori gauge fixing, the light-cone gauge fixing
\cite{Goroff} are just a few of them.

The main purpose of our work is to aboard the Dirac's analysis for
Einstein's linearized gravity by making some analogies with
Maxwell's theory as the analogous Coulomb's gauge fixing for
gravity. A comparison with the Maskawa-Nakajima\cite{MN} procedure
is also discussed. As usual, we'll decompose tensors of rank 1 and
2 following the known methodology\cite{Xiang}.

This paper is organized as follows. In the next section we study
Dirac's procedure for Einstein's Linearized Gravity with the
Coulomb's gauge. Then, we perform the Maskawa-Nakajima reduction
to compare Maxwell's theory with Linearized Gravity in the
previously mentioned gauge, where the projectors of spin 1 and 2
merge naturally. Finally, we end with some comments.

\section{Notation}

The Hilbert-Einstein action, $S_{HE}$ describes gravity under the
postulates of general relativity and it comes as
\begin{equation}
S_{HE} = -\frac{1}{k^{N-2}} \int d^{N-1}x \sqrt{-g}
\mathcal{R}\,\,, \label{1}
\end{equation}
where $g$ is the determinant of the metric tensor, $k$ is a
proportionality constant which comes in units of length and
$\mathcal{R}$ is the Ricci's scalar, defined as
\begin{equation}
\mathcal{R}  = g^{\mu\nu} \mathcal{R}_{\mu\nu}  = g^{\mu\nu}
{\mathcal{R}^{\lambda}}_{\mu\lambda\nu}\,\,, \label{2}
\end{equation}
where $\mathcal{R}_{\mu\nu}$ are the components of Ricci's tensor
and ${\mathcal{R}^{\lambda}}_{\mu\sigma\nu}$ the components of
 Riemman-Christoffel's tensor. Here we think on a N-dimensional
($N>2$) space-time with null metricity ($\nabla_{\alpha}
g_{\mu\nu} = 0$) and torsionless (${T^{\alpha}}_{\mu\nu} = 0$)
therefore, the Levi-Civita's connection comes from Christoffel's
symbols as ${\Gamma^{\mu}}_{\lambda\nu}$ only in terms of the
components of the metric tensor and its first derivatives in the
usual way ${\Gamma^{\mu}}_{\lambda\nu} =
\frac{g^{\mu\rho}}{2}(\partial_{\lambda}g_{\rho\nu} +
\partial_{\nu}g_{\rho\lambda} - \partial_{\rho}g_{\lambda\nu})$.

So, the components of Riemann-Christoffel's tensor comes as
\begin{align}
{\mathcal{R}^{\alpha}}_{\mu\nu\lambda} = \partial_{\lambda}
{\Gamma^{\alpha}}_{\nu\mu} - \partial_{\nu}
{\Gamma^{\alpha}}_{\lambda\mu} + {\Gamma^{\alpha}}_{\lambda\sigma}
{\Gamma^{\sigma}}_{\nu\mu} - {\Gamma^{\alpha}}_{\nu\sigma}
{\Gamma^{\sigma}}_{\lambda\mu}\,\,, \label{3}
\end{align}
establishing that Ricci's scalar in the action (\ref{1}) have a
dependence until second order in derivatives of the components of
the metric.

When one perform arbitrary functional variations on the metric at
the Hilbert-Einstein's action, it can be shown that it's an
extremal if
\begin{equation}
G_{\alpha\beta}\equiv \mathcal{R}_{\alpha\beta} -
\frac{g_{\alpha\beta}}{2}\mathcal{R} = 0 \,\,, \label{4}
\end{equation}
where $G_{\alpha\beta}$ is the Einstein's tensor. With all these,
the perturbative analysis is thought in the surrounding
stationary points of the action $S_{HE}$.

First order perturbations in the metric are made in a usual way
around a Minkowski's background, this means
\begin{align}
g_{\mu\nu} = \eta_{\mu\nu} + h_{\mu\nu}\,\,, \label{5}
\end{align}
\begin{align}
g^{\mu\nu} &= \eta^{\mu\nu} - h^{\mu\nu}\,\,, \label{6}
\end{align}
where $\eta = diag(-1,1,...,1)$ is the Minkowski's metric tensor
and $h_{\mu\nu} << 1$ is the perturbation. At first order, we rise
and down indexes with Minkowski's metric.

Now, we can write the linearized version of (\ref{1}) in terms
of the field $h_{\mu\nu}$ as follows
\begin{equation}
S^{L} = -\frac{1}{k^{N-2}} \int d^{N-1}x  h_{\mu\nu} G^{L
\mu\nu}(h) \,\,, \label{7}
\end{equation}
where
\begin{align}
G^{L}_{\mu\nu}(h) &= \mathcal{R}^{L}_{\mu\nu} -
\frac{\eta_{\mu\nu}}{2}\mathcal{R}^{L}\nonumber \\
&=\frac{1}{2}[\square h_{\mu\nu} +
\partial_{\mu}\partial_{\nu}h - \partial_{\mu}\partial_{\alpha}h^{\alpha}_{\nu}\nonumber \\
&-\partial_{\nu}\partial_{\alpha}h^{\alpha}_{\mu} +
\eta_{\mu\nu}(\square h -
\partial_{\alpha}\partial_{\beta}h^{\alpha\beta})]\,\,. \label{8}
\end{align}

In this context, the perturbative field $h_{\mu\nu}$ is a rank 2
tensor that transforms under the (locally) Lorentz group and, due
to the diffeomorphism's symmetry, it also has a the well known
functional transformation
\begin{align} \delta h_{\mu\nu}=
\partial_\mu \xi_\nu + \partial_\nu \xi_\mu\,\,, \label{8a}
\end{align}
where $\xi_\mu $ is an arbitrary vector field, which is continuous
and differentiable, recalling us in some way to the form in which
the vector potential of Maxwell's theory transforms due to the
gradient of a scalar field. It's very common to find references of
this diffeomorphism transformation as a gauge invariance that can
be found formally using Castellani's procedure \cite{Castellani}.
In this sense, we'll use this to make the transition to the second
class constraints system, i.e. fixing a gauge in linearized
gravity having in mind the analogy with the fixing of Coulomb's
gauge.

\section{Dirac's procedure in the perturbative regime}
We begin by doing a $(N-1)+1$ decomposition of the linearized
action (\ref{7}) that we have redefined as $S^{*L} \equiv
-k^{N-2}S^{L}$, so it follows
\begin{align}
S^{*L} &= \bigg\langle \frac{1}{2}\dot{h}_{ij}\dot{h}_{ij} - \frac{1}{2}\bigg(\dot{h}_{kk}\bigg)^2 +
2\bigg(\dot{h}_{kk} \partial_l h_{0l} - \dot{h}_{kl} \partial_k h_{0l}\bigg) \nonumber \\
&+ h_{0l}\bigg( - \bigtriangleup h_{0l} + \partial_l \partial_k h_{0k}\bigg)
+ h_{ij}\bigg(\frac{1}{2}\bigtriangleup h_{ij} - \partial_i \partial_k h_{kj} \nonumber \\
&+ \partial_i \partial_j h_{kk} -
\frac{\eta_{ij}}{2}\bigtriangleup h_{kk}\bigg) +
h_{00}\bigg(\bigtriangleup h_{kk} - \partial_l \partial_k
h_{kl}\bigg) \bigg\rangle  \,\,. \label{9}
\end{align}

Following the Dirac's procedure, we write the canonical momentum
$\Pi^{\alpha \beta} \equiv \frac{\partial L}{\partial
\dot{h}_{\alpha \beta}}$, to get
\begin{align}
\Pi^{0\mu} &= 0 \,\,, \label{momento1}
\end{align}
\begin{align}
\Pi^{ij} &= \dot{h}_{ij} - \eta_{ij}\dot{h}_{kk} + 2 \eta_{ij}
\partial_l h_{0l} - \partial_i h_{0j} - \partial_j h_{0i} \,\,.
\label{momento2}
\end{align}

We notice that (\ref{momento1}) is a primary constraint
$\phi_{1}^{\mu} \equiv \Pi^{0\mu}$, while (\ref{momento2}) is an
expression that allow us to find the velocities $\dot{h}_{ij}$.
With these, we can write the Hamiltonian density $H_{0}$ of the
system
\begin{align}
H_{0} &= \frac{\Pi_{ij}\Pi_{ij}}{2} - \frac{(\Pi_{kk})^2}{2(N-2)} - h_{ij}
\bigg(\frac{\bigtriangleup h_{ij}}{2} - \eta_{ij} \bigtriangleup h_{kk} \nonumber \\
&- \partial_i \partial_k h_{kj} - \partial_j \partial_k h_{ki} + \partial_i \partial_j h_{kk} \bigg)
+ h_{00} \bigg( \partial_l \partial_k h_{kl} \nonumber \\
&- \bigtriangleup h_{kk} \bigg) - 2h_{0j} \partial_i \Pi^{ij}
\,\,,\label{GH0}
\end{align}
and the total Hamiltonian density can be build if we include the
primary constraint with a Lagrange multiplier
\begin{align}
H_{T} = H_{0} + u_{\mu} \Pi^{0\mu} \,\,.\label{10}
\end{align}

To continue with the procedure, we preserve the primary constraint
${\phi_1}^\mu $ using Poisson's brackets algebra for symmetric
rank 2 fields, which by construction they come as:
\begin{align}
\bigg\lbrace h_{\alpha \beta}(x), \Pi^{\mu \nu}(y) \bigg\rbrace =
\frac{1}{2} ( \delta_{\alpha}^{\mu} \delta_{\beta}^{\nu} +
\delta_{\alpha}^{\nu} \delta_{\beta}^{\mu} )
\delta^{N-1}(x-y)\,\,.\label{11}
\end{align}
Hence, the preservation of the aforementioned constraint gives
\begin{align}
\dot{\phi}_{1}^{\mu}(x) &= \int d^{N-1}y \bigg\lbrace \phi_{1}^{\mu}(x), H_{T}(y) \bigg\rbrace  \nonumber \\
&= \delta^{\mu}_{0}(\bigtriangleup h_{kk} - \partial_l \partial_k
h_{lk}) + \delta^{\mu}_{j} \partial_i \Pi^{ij} = 0\,\,,\label{12}
\end{align}
representing $N$ new constraints which components are $\phi_{2}^{0}
\equiv \bigtriangleup h_{kk} - \partial_l \partial_k h_{lk}$ and
$\phi_{2}^{j} \equiv \partial_i \Pi^{ij}$. We must preserve these,
so it can be obtained
\begin{align}
\dot{\phi}_{2}^{0} = \int d^{N-1}y \bigg\lbrace \phi_{2}^{0}(x),
H_{T}(y) \bigg\rbrace = -\partial_i \phi_{2}^{i} \equiv 0
\,\,,\label{14}
\end{align}
\begin{align}
\dot{\phi}_{2}^{j} = \int d^{N-1}y \bigg\lbrace \phi_{2}^{j}(x),
H_{T}(y) \bigg\rbrace = 0\,\,.\label{15}
\end{align}

No more new constraints appear, so the process of preservation
ends. Therefore, we resume the constraints
\begin{align}
\phi_{1}^{\mu} &\equiv \Pi^{0\mu}\,\,, \label{vinculo1}
\end{align}
\begin{align}
\phi_{2}^{0} &\equiv \bigtriangleup h_{kk} - \partial_l \partial_k
h_{lk} \,\,,\label{vinculo2}
\end{align}
\begin{align}
\phi_{2}^{j} &\equiv \partial_i \Pi^{ij}\,\,, \label{vinculo3}
\end{align}
and we immediately note that all of them are first class
constraints. Ahead we'll extend this system to a second class one
when we choose a gauge. The physical reason of this comes from the
ambiguity due to the gauge freedom that lead us to the fact that
not all of the fields are actually local degrees of freedom. This
is confirmed by noticing that the Hamiltonian (\ref{GH0}) is not
positively defined in analogy with Maxwell's theory with gauge
freedom.

Considering then that Einstein's linearized gravity have an $N$
parameters gauge invariance presented in (\ref{8a}), we choose $N$
gauges similar to the Coulomb gauge via the following ad hoc
constraints
\begin{equation}
\chi_\mu = \partial_i h_{i\mu}\,\,, \label{16}
\end{equation}
meaning $N$ new constraints impossed which must be preserved as
the Dirac's procedure says. So the preservation of them leads
to
\begin{align}
\dot{\chi}_{\mu} &= \int d^{N-1}y \bigg\lbrace \chi_{\mu}(x), H_{T}(y) \bigg\rbrace \nonumber \\
&= \int d^{N-1}y \bigg\lbrace \chi_{\mu}(x), H_{0}(y) \bigg\rbrace
+ \frac{1}{2}\delta_{\mu}^{0} \partial_i u_{i}(x)\,\,.
\label{chimu}
\end{align}
If we take $\mu = 0$, we get a differential equation for $N-1$
Lagrange multipliers without getting any new constraints, this
means
\begin{align}
\partial_i u_{i}(x) \simeq - 2 \int
d^{N-1}y \bigg\lbrace \partial_i h_{i0}(x), H_{0}(y)
\bigg\rbrace\,\,.\label{17}
\end{align}

However, if we take $\mu = j$ in (\ref{chimu}) we get $N-1$ new constraints
\begin{align}
\chi_{j}^{2} \equiv -\frac{1}{N-2} \partial_j \Pi_{kk} +
\bigtriangleup h_{0j}\,\,,\label{18}
\end{align}
and their preservation give
\begin{align}
\dot{\chi}_{j}^{2} = \int d^{N-1}y \bigg\lbrace \chi_{j}^{2}(x), H_{0}(y) \bigg\rbrace + \frac{1}{2} \bigtriangleup u_{j}(x)  \label{chi2}
\end{align}
which means $N-1$ Poisson type equations for the multipliers
$u_{j}$ and due to consistency with (\ref{17}) and using
(\ref{18}) they can be written as follows
\begin{align}
\partial_j \dot{\chi}_{j}^{2} &= \partial_j \int d^{N-1}y \bigg\lbrace \chi_{j}^{2}(x), H_{0}(y) \bigg\rbrace +
\frac{1}{2} \bigtriangleup \partial_j u_{j}(x) \nonumber \\
&= - \frac{1}{N-2} \bigtriangleup \int d^{N-1}y \bigg\lbrace
\Pi_{kk}(x), H_{0}(y) \bigg\rbrace \nonumber \\
&=(N-2)\bigtriangleup^{2} h_{00}(x) \,\,, \label{19}
\end{align}
which up to harmonic forms we get a new constraint
\begin{equation}
\chi_{3} \equiv h_{00} \,\, .\label{20}
\end{equation}

Its preservation follows as
\begin{align}
\dot{\chi}_{3} = \int d^{N-1}y \bigg\lbrace \chi_{3}(x), H_{T}(y)
\bigg\rbrace = u_{0}(x) \,\,,\label{21}
\end{align}
and with this we can determine the remaining Lagrange multiplier
and the preservation process ends.

We rename the constraints and make a list of all of them in the following way
\begin{align}
\chi_{1} &\equiv \Pi^{00} \,\,,\label{vinc1}
\end{align}
\begin{align}
\chi_{1}^{i} &\equiv \Pi^{0i} \,\,,\label{vinc2}
\end{align}
\begin{align}
\chi_{2} &\equiv h_{kk} \,\,,\label{vinc3}
\end{align}
\begin{align}
\chi_{2}^{j} &\equiv \partial_i \Pi^{ij}\,\,, \label{vinc4}
\end{align}
\begin{align}
\chi_{3}^{j} &\equiv \partial_i h_{ij}\,\,, \label{vinc5}
\end{align}
\begin{align}
\chi_{4} &\equiv \Pi_{kk}\,\,, \label{vinc6}
\end{align}
\begin{align}
\chi_{4}^{j} &\equiv h_{0j}\,\,, \label{vinc7}
\end{align}
\begin{align}
\chi_{5} &\equiv h_{00} \,\,.\label{vinc8}
\end{align}
This means that we have a system of $4N$ second class constraints,
and since there are $N(N+1)$ fields and canonically conjugate
momenta, we finally have $\frac{N(N+1)-4N}{2} = \frac{N(N-3)}{2}$
degrees of freedom.

The next step is to build Dirac's matrix using the Poisson
brackets of the constraints, so we get a $4N\times4N$ range
matrix that is written as
\begin{equation}
\scalemath{0.75}{
C = \begin{pmatrix}
0 & 0 & 0 & 0 & 0 & 0 & 0 & -1 \\
0 & 0 & 0 & 0 & 0 & 0 & -\frac{\eta_{ij}}{2} & 0 \\
0 & 0 & 0 & -\partial_i  & 0 & (N-1) & 0 & 0 \\
0 & 0 & \partial_j  & 0 & \frac{\eta_{ij}\bigtriangleup + \partial_i \partial_j }{2}  & 0 & 0 & 0 \\
0 & 0 & 0 & - \frac{\eta_{ij}\bigtriangleup + \partial_i \partial_j }{2}  & 0 & \partial_j  & 0 & 0 \\
0 & 0 & -(N-1) & 0 & -\partial_i  & 0 & 0 & 0 \\
0 & \frac{\eta_{ij}}{2} & 0 & 0 & 0 & 0 & 0 & 0 \\
1 & 0 & 0 & 0 & 0 & 0 & 0 & 0 \\
\end{pmatrix}
\delta^{N-1}_{(x-x')}}  \,\,.\label{matriz}
\end{equation}

To continue Dirac's procedure it is necessary to find the inverse
$ C^{-1}(x'-y)$, which must satisfies the property
\begin{equation}
\int d^{N-1}x'  C(x-x') C^{-1}(x'-y) = \mathbb{I}
\delta^{N-1}(x-y)  \,\,, \label{propiedad}
\end{equation}
where $\mathbb{I}$ is the identity matrix with $4N\times4N$ range.

We make an ansatz over the form of this inverse matrix, in a similar
way to the form that the original matrix $ C(x-x')$ have, to get
\begin{equation}
\scalemath{0.75}{
C^{-1} = \begin{pmatrix}
0 & 0 & 0 & 0 & 0 & 0 & 0 & \delta^{N-1} \\
0 & 0 & 0 & 0 & 0 & 0 & 2 \eta_{jl} \delta^{N-1} & 0 \\
0 & 0 & 0 & -A_{j}  & 0 & -E & 0 & 0 \\
0 & 0 & A_{l}  & 0 & -B_{jl}  & 0 & 0 & 0 \\
0 & 0 & 0 & B_{jl}  & 0 & -D_{l}  & 0 & 0 \\
0 & 0 & E & 0 & D_{j}  & 0 & 0 & 0 \\
0 & -2 \eta_{jl} \delta^{N-1} & 0 & 0 & 0 & 0 & 0 & 0 \\
-\delta^{N-1} & 0 & 0 & 0 & 0 & 0 & 0 & 0 \\
\end{pmatrix}_{(x'-y)}}  \,\,,\label{matrizinv}
\end{equation}
where $A_{i}(x'-y)$, $B_{ij}(x'-y)$, $D_{i}(x'-y)$ and $E(x'-y)$
are undetermined functions. From (\ref{propiedad}) it arises a set
of consistency rules
\begin{align}
- \partial_i A_i (x-y) + (N-1)C(x-y) &= \delta^{N-1}_{(x-y)} \label{siseqi}\\
\partial_j B_{ij}(x-y) + (N-1)D_i (x-y) &= 0 \\
- \partial_j A_i (x-y) + \bigg( \frac{\eta_{lj}\bigtriangleup + \partial_l \partial_j}{2} \bigg) B_{lj}(x-y) &= \eta_{ij} \delta^{N-1}_{(x-y)} \\
\partial_j C(x-y) + \bigg( \frac{\eta_{ij}\bigtriangleup + \partial_i \partial_j}{2} \bigg) D_i (x-y) &= 0 \\
- \bigg( \frac{\eta_{ij}\bigtriangleup + \partial_i \partial_j}{2} \bigg) A_j (x-y) + \partial_i C(x-y) &= 0 \\
\bigg( \frac{\eta_{lj}\bigtriangleup + \partial_l \partial_j}{2} \bigg) B_{il} (x-y) + \partial_j D_i (x-y) &= \eta_{ij} \delta^{N-1}_{(x-y)} \\
(N-1) A_i (x-y) - \partial_j B_{ij} (x-y) &= 0 \\
(N-1) C(x-y) + \partial_j D_j (x-y) &= \delta^{N-1}_{(x-y)} \,\,,
\label{siseqf}
\end{align}
which allow us to determine the unknown functions in the inverse
matrix. For this task we use the solution of the $N>2$ dimension
Poisson's equation \cite{Poisson}, so we find a solution of the
system (\ref{siseqi} - \ref{siseqf}) as follows

\begin{align}
A_i (x-y) = - D_i (x-y)\,\,, \label{inv1}
\end{align}
\begin{align}
B_{ij} (x-y) =  \bigg(2 \eta_{ij}
+\frac{(N-3)}{(N-2)}\hat{\partial}_i \hat{\partial}_j
\bigg)\phi(x-y)\,\,, \label{inv2}
\end{align}
\begin{align}
C(x-y) = \frac{1}{N-2}\delta^{N-1}(x-y)\,\,, \label{inv3}
\end{align}
\begin{align}
D_j (x-y) = -\frac{1}{(N-2)}\partial_j\phi(x-y)\,\,, \label{inv4}
\end{align}
where
\begin{equation}
\phi(x-y) \equiv \left\{
    \begin{array}{ll}
        \frac{1}{2\pi}ln\vert x-y\vert  & \mbox{if } N = 3 \\
        -\frac{1}{\alpha_{(N)} \vert x-y\vert^{N-3}} & \mbox{if } N \geq 4
    \end{array}
\right. \,\,, \label{poisson}
\end{equation}
with $\alpha_{(N)} =
\frac{\Gamma(\frac{N-1}{2})}{2(N-3)\pi^{\frac{N-1}{2}}}$ denotes
the surface area of a $N$-sphere in $\mathbb{R}^{N}$ and
$\hat{\partial}_i\equiv \frac{\partial_i}{\sqrt{-\Delta}}$.

With this, we build Dirac's brackets in the usual way $\lbrace
A(x'), B(y') \rbrace_D = \lbrace A(x'), B(y') \rbrace - \int
d^{N-1}x d^{N-1}y \lbrace A(x'), \chi_{s}(x) \rbrace
C^{-1}_{ss'}(x-y) \lbrace \chi_{s'}(y), B(y') \rbrace$ and all the
constraints are now first class, so we can take them strongly
equal to zero. So the brackets that are not zero are
\begin{align}
&\lbrace h_{ij}(x), \Pi^{kl}(y) \rbrace_D \nonumber \\
&= \bigg[\frac{1}{2}(\eta_{ik} \eta_{jl} + \eta_{il} \eta_{jk} + \eta_{ik} \hat{\partial}_j
\hat{\partial}_l + \eta_{jl} \hat{\partial}_i \hat{\partial}_k \nonumber \\
&+ \eta_{il} \hat{\partial}_j \hat{\partial}_k + \eta_{jk} \hat{\partial}_i \hat{\partial}_l )
- \frac{1}{N-2}(\eta_{ij} \eta_{kl} + \eta_{ij} \hat{\partial}_k \hat{\partial}_l \nonumber \\
&+ \eta_{kl} \hat{\partial}_i \hat{\partial}_j ) +
\frac{N-3}{N-2}\hat{\partial}_i \hat{\partial}_j \hat{\partial}_k
\hat{\partial}_l \bigg] \delta^{N-1}(x-y) \,\,.
\label{corchetedirac}
\end{align}

\section{Maskawa-Nakajima's analysis}

\subsection{Maxwell Field}

Before to explore the Maskawa-Nakajima's (MN) reduction for
Einstein's linearized gravity, we shall do a brief and pedagogical
review of the reduction for Maxwell's electromagnetic theory for a
better understanding of some of the aspects that we want to point
out.

Maxwell's theory is described by the action
\begin{equation}
S = \bigg\langle \frac{1}{4} F_{\mu \nu} F^{\mu \nu} \bigg\rangle
\,\,, \label{accionMaxwell}
\end{equation}
where $F_{\mu\nu} = \partial_{\mu} A_{\nu} - \partial_{\nu}
A_{\mu}$ is the Maxwell tensor and $A_{\mu}$ is the potential
field.

Maxwell's theory is invariant under Lorentz group and the gauge
group $U(1)$, hence there are redundant degrees of freedom so the
theory must be reducible. A possible way to write down the reduced
action $\bar{S}$ starts with a standard decomposition of the
potential field in it's transverse and longitudinal parts,
following the known prescription
$A_{i}={A^T}_{i}+\partial_i{A^L}$. By doing this, we can eliminate
the fields that are not dynamical to rewrite the action as
\begin{equation}
\bar{S} = \bigg\langle \frac{1}{2} {A^{T}}_{i} \square {A^{T}}_{i}
\bigg\rangle \,\,,\label{accionMaxwellreduc}
\end{equation}
it can be noted the transverse part of the field $A_{i}$ as the
only field which carry the physical propagation.

So, the analysis of Lagrangian constraints of Maxwell's theory
tells us that the temporal component of the field, in other words
$A_0$ does not propagate, which is supported by the fact that this
component appears as the Lagrange multiplier associated to the
Gauss constraint in the Hamiltonian formalism and also it's
canonical conjugate momentum $\Pi_0$ is a primary constraint. This
allows us to focus our attention on the sub-space spanned by the
$N-1$ purely spacial components of the potential field, this means
$A_i$, whether it's in the configurations or in the phase space.
The $(N-1)$ spacial part of the configuration space have $2(N-1)$
dimension because we haven't chosen a gauge yet, and we name this
space as $\varepsilon_{(A_i,\dot{A}_i)}$. However, the $(N-1)$
spacial section of the phase space denoted by
${\varepsilon}^*_{(A_i,\Pi_i)}$ is not necessarily locally
isomorph to $\varepsilon_{(A_i,\dot{A}_i)}$, which only relies
formally on whether or not we impose the Gauss constraint from the
beginning.

So, starting with ${\varepsilon}^*_{(A_i,\Pi_i)}$ and performing
the gauge fixing (i.e., the Coulomb gauge) and the Gauss
constraint mean two constraints on the $(N-1)$ spacial section of
the phase space which conduce to a new and reduced one, which we
shall call $\bar{{\varepsilon}^*}_{(\bar{A}_i,\bar{\Pi}_i)}$ with
dimension $2(N-2)$.

MN theorem\cite{MN} tells us that there exist a non one to one
map from the space ${\varepsilon}^*_{(A_i,\Pi_i)}$ into the
new space $\bar{{\varepsilon}^*}_{(\bar{A}_i,\bar{\Pi}_i)}$ which
represents the physical degrees of freedom reduction and it
must have consistency with the brackets obtained via the Dirac's
reduction procedure. In this sense, we assume then a matrix
representation of the projection, $\Omega$ for the $N-1$ spacial
components as follows
\begin{align}
\bar{A}_{k} = \Omega_{kl}A_{l}\,\,,\label{mapa1}
\end{align}
\begin{align}
\bar{\Pi}_{k} = \Omega_{kl}\Pi_{l}\,\,,\label{mapa2}
\end{align}
where we expect that (\ref{mapa2}) is redundant due to the Gauss
constraint $\bar{\Pi}_{i} = \Pi_{i}$.

This projection applied to the Poisson brackets of the fields in
$\varepsilon_{(A,\Pi)}$, this means $\{A_{k}(x) , \Pi_{l}(y) \} =
\eta_{kl} \delta^{N-1}(x-y)$ leads us to
\begin{align}
\{\bar{A}_{k}(x) , \bar{\Pi}_{l}(y) \} =
\Omega_{km}(x)\Omega_{lm}(y)
\delta^{N-1}(x-y)\,\,.\label{corcheteProy}
\end{align}

We haven't said anything about the form of $\Omega$, but since
$\bar{A}_k$ are transverse fields, we realize this transformation
via the $N-1$ transverse projector invariant under parity in the
following form
\begin{equation} \Omega_{ij}(x) = \Omega_{ij}(-x) \equiv
\eta_{ij} + \hat{\partial}_i \hat{\partial}_j\,\,,\label{ProyS1}
\end{equation}
which satisfy
\begin{equation}
\Omega_{km}\Omega_{lm}=\Omega_{kl}\,\,.\label{PropProyS1}
\end{equation}

Also, we can verify with this projector that the Poisson brackets
that are defined in $\varepsilon_{(A_i,\Pi_i)}$ induce the
expected form of Dirac's brackets \cite{HRT} in
$\bar{\varepsilon}_{(\bar{A}_i,\bar{\Pi}_i)}$.

The advantage of this method is that, beyond the $2+1$ dimensional
case, it would be a not easy bussines to find the explicit and
irreducible decomposition of any tensor field of arbitrary rank
and therefore the writing of Dirac's brackets in the reduced space
following the MN procedure.

\subsection{Linearized gravity}

Taking in mind the last discussion, now we want to follow a
similar trail to a MN reduction for Einstein's linearized gravity.
We begin by making an ADM\cite{ADM} decomposition in (\ref{9}),
exposing the transverse (T), longitudinal (L) and
traceless-tranverse (Tt) parts in the way $h_{ij} = h_{ij}^{Tt} +
h_{ij}^{T} + h_{ij}^{L}$. After this, the non dynamical fields can
be removed and the reduced action is
\begin{equation}
\bar{S} = \bigg\langle \frac{1}{2} {h^{Tt}}_{ij} \square
{h^{Tt}}_{ij} \bigg\rangle \,\,,\label{GravRed}
\end{equation}
which clearly shows that only the Tt part of the field propagates
degrees of freedom.

From the phase space point of view, to exhibit the reduction of a
phase sub-space, this means $\varepsilon^*_{(h_{ij},\Pi_{ij})}$ in
to other $\bar{\varepsilon^*}_{(\bar{h}_{ij},\bar{\Pi}_{ij})}$
with dimension $N(N-3)$, we must consider the $N$ gauge fixings
provided in (\ref{16}) and the $N$ Gauss constraints rewritten as
$\phi_{2}^{\mu} \equiv
\partial_i \Pi^{i\mu}$ with the help of the primary constraints
(\ref{vinculo1}). Then, we realize this reduction through
traceless-transverse projector as
\begin{align}
\bar{h}_{ij} = \Omega_{ijmn}h_{mn} \,\,,\label{Gravproy1}
\end{align}
\begin{align}
\bar{\Pi}_{ij} = \Omega_{ijmn}\Pi_{mn} \,\,,\label{Gravproy2}
\end{align}
where the following algebraic properties must be satisfied
$\Omega_{ijmn} = \Omega_{jimn} = \Omega_{ijnm}$ and
\begin{align}
\Omega_{ijkl}\Omega_{mnkl} = \Omega_{ijmn}\,\,, \label{prop1}
\end{align}
\begin{align}
\partial_i \bar{h}_{ij} = \Omega_{ijmn} \partial_i h_{mn} = 0\,\,,\label{prop2}
\end{align}
\begin{align}
\bar{h}_{ii} = \Omega_{iimn} h_{mn} = 0 \,\,. \label{prop3}
\end{align}

An ansatz on the form of $\Omega_{ijmn}$ is
\begin{equation}
\Omega_{ijmn} = \frac{\alpha(N)}{2}( \Omega_{im}\Omega_{jn} +
\Omega_{in}\Omega_{jm} ) + \frac{\beta(N)}{2}
\Omega_{ij}\Omega_{mn}\,\,, \label{ansatz}
\end{equation}
where $\alpha(N)$ and $\beta(N)$ are unknown real coefficients and
with the help of (\ref{prop1} - \ref{prop3}) we can find them as
\begin{align}
\alpha(N) = 1\,\,, \label{alfa}
\end{align}
\begin{align}
\beta(N) = - \frac{2}{N-2}\,\,, \label{beta}
\end{align}
therefore, the projector take the form
\begin{equation}
\Omega_{ijkl} = \frac{1}{2}( \Omega_{ik}\Omega_{jl} +
\Omega_{il}\Omega_{jk} ) - \frac{1}{N-2}
\Omega_{ij}\Omega_{kl}\,\,. \label{proyFinal}
\end{equation}

This projector is applied to the Poisson brackets of the fields in
$\varepsilon^*_{(h_{ij},\Pi_{ij})}$, in other words $\{ h_{ij}(x)
, \Pi_{kl}(y) \} = \frac{(\eta_{ik}\eta_{jl} +
\eta_{il}\eta_{jk})}{2} \delta^{N-1}(x-y)$, and this gives us
\begin{align}
\{\bar{h}_{ij}(x) , \bar{\Pi}_{kl}(y) \} =
\Omega_{ijmn}(x)\Omega_{klmn}(y) \delta^{N-1}(x-y)\,\,,
\label{corcheteFinal}
\end{align}
which is equivalent to Dirac's bracket (\ref{corchetedirac}).

\section{Conclusion}

The symmetries of physical systems imply the existence of
conserved quantities according to Noether's theorem. This is so
that when we study the action of a given system where translation
invariance induces the conservation in the lineal momentum, the
conservation of energy comes from the invariance under time
displacements, and so on. But all of this is accompanied by the
fact that the fields that describe the theory can't be written in
a unique way. Whether it is because they change under certain
coordinates transformation groups or because they can
simultaneously transform under functional variations as well. A
typical case of this is the Maxwell's electromagnetic theory
described by a tensor field of rank 1 that transforms under the
Lorentz Group, which in $N$ dimensions is denoted as $ISO(N-1,1)$,
and under the $U(1)$ group that represents the gauge invariance.

In this sense, Einstein's linearized gravity is very similar to
Maxwell's case. In the perturbative regime, the symmetrical rank 2
tensor field transforms under the (local) Lorentz group and also
functionally under a diffeomorphism which would be thought as a
gauge transformation.

No matter which system with symmetries we're studying, there's not
a unique way to approach to the true configuration of the physical
fields that describe the system. From the configuration space's
perspective, it is possible to make the reduction following the
analysis of Lagrangian constraints. However, in this work we've
focused in the phase space and the reduction via Dirac's and MN
procedures, laying down the groundwork for a possible quantization
of linearized gravity.

Then, we applied Dirac's procedure to the Einstein's linearized
gravity to find the correct algebra for the minimum physical
fields of the theory in agreement with reference\citep{EvensK}.
Particularly, we choose a Coulomb gauge to assure the minimal
number of degrees of freedom, in analogy with the Maxwell's case.

Finally, the MN reduction has been performed conjecturing that is
possible to project the unconstrained Poisson brackets on to the
reduced and the algebra obtained is consistent with Dirac's
procedure thanks to the use of the $N(N-1)$ traceless-tranverse
projector for the rank 2 tensor fields, in the same way that it's
used with the $N-1$ tranverse projector for the rank 1 tensor
field in Maxwell's theory.

If we analyze this in perspective with all the analogies that
exist between Einstein's linearized gravity and Maxwell theory, it
would be interesting to explore a first order formalism for the
theory, where we can confirm the reduction of the degrees of
freedom, the gauge invariance, and the brackets, but this is the
topic for a future work.

\end{document}